\pdfoutput=1
\documentclass[12pt]{article}

\font\grande=cmr9.5 scaled \magstep4
\font\medio=cmr9.5 scaled \magstep2
\outer\def\beginsection#1\par{\medbreak\bigskip
      \message{#1}\leftline{\bf#1}\nobreak\medskip
\vskip-\parskip
      \noindent}
\usepackage{graphicx} 
\begin{document}
\bibliographystyle {unsrt}

\titlepage

\begin{flushright}
CERN-PH-TH/2015-195
\end{flushright}

\vspace{10mm}
\begin{center}
{\grande Hypermagnetic gyrotropy, inflation}\\
\vspace{5mm}
{\grande and the baryon asymmetry of the Universe}\\
\vspace{1.5cm}
 Massimo Giovannini
 \footnote{Electronic address: massimo.giovannini@cern.ch}\\
\vspace{1cm}
{{\sl Theory Division, CERN, 1211 Geneva 23, Switzerland }}\\
\vspace{0.5cm}
{{\sl INFN, Section of Milan-Bicocca, 20126 Milan, Italy}}
\vspace*{0.5cm}
\end{center}

\vskip 0.5cm
\centerline{\medio  Abstract}
We investigate the production of the hypermagnetic gyrotropy  
when the electric and magnetic gauge couplings evolve at different rates, as it happens 
in the the relativistic theory of the Van der Waals forces. If a pseudo-scalar interaction breaks the duality symmetry of the corresponding equations, the gyrotropic configurations of the hypermagnetic fields can be amplified from the vacuum during an inflationary stage of expansion. After charting the parameter space of the model in terms of the rates of evolution of the magnetic and electric gauge couplings, we identify the regions where the gyrotropy is sufficiently intense to seed the baryon asymmetry of the Universe at the electroweak epoch while the backreaction constraints, the strong coupling bounds and the other astrophysical limits are concurrently satisfied. 
\vskip 0.5cm

\noindent

\vspace{5mm}

\vfill
\newpage
Since the seminal work of Sakharov \cite{sak} various attempts have been made to account for the baryon asymmetry 
of the Universe (BAU in what follows). The standard lore of baryogenesis (see e.g. the first article of Ref. \cite{CKN}) stipulates that during a strongly 
first-order electroweak phase transition the expanding bubbles are nucleated while the baryon number 
is violated by sphaleron processes. Given the current value of the Higgs mass, to produce a sufficiently strong (first-order) phase transition and to get enough $CP$ violation at the 
bubble wall, the standard electroweak theory must be appropriately extended. The second complementary lore for the generation of the BAU is leptogenesis (see e.g. the second article of Ref. \cite{CKN}) which can be conventionally realized thanks to heavy Majorana neutrinos decaying out of equilibrium and producing an excess of lepton 
number ($L$ in what follows). The excess in $L$ can lead to the observed baryon number thanks to sphaleron interactions violating $(B+ L)$. 
An admittedly less conventional perspective stipulates that the BAU could be the result of the decay of maximally helical configurations of the hypercharge field (sometimes dubbed hypermagnetic knots) \cite{mg1}. Indeed, while the $SU_{L}(2)$ anomaly is typically responsible for $B$ and $L$ non-conservation via instantons and sphalerons, the $U_{Y}(1)$ anomaly might lead to the transformation 
of the infra-red modes of the hypercharge field into fermions \cite{rub}. As previously suggested (see third and fourth paper of \cite{mg1}) the production of the BAU demands, in this 
context, the dynamical generation of the hypermagnetic gyrotropy\footnote{The magnetic and kinetic gyrotropies 
play a crucial role in the mean-field dynamo \cite{vain} and the same notion occurs, with the due differences, in the present context.} ${\mathcal G}^{(B)}(\vec{x}, \tau) = \vec{B}_{Y}\cdot \vec{\nabla} \times \vec{B}_{Y}$  (where $\vec{B}_{Y}$ denotes the hypermagnetic field).  
This hypothesis has been subsequently scrutinized, within diverse frameworks, by several authors (see e.g. \cite{sev} for an incomplete list of references). 

In this investigation we propose a specific scenario accounting at once for the formation 
of the hypermagnetic knots and for the existence of large-scale magnetic. 
We shall therefore posit a derivative interaction of the hypercharge field with one or more scalar fields
during a quasi-de Sitter stage of expansion. Let us consider, fo the sake of concreteness the following four-dimensional action:
\begin{equation}
S = \int d^4 x \, \sqrt{- g} \biggl[ {\mathcal M}_{\sigma}^{\rho}(\varphi,\psi) 
Y_{\rho\alpha}\, Y^{\sigma\alpha} - {\mathcal N}_{\sigma}^{\rho}(\varphi,\psi) \widetilde{Y}_{\rho\alpha}\, \widetilde{Y}^{\sigma\alpha}
+ {\mathcal Q} _{\sigma}^{\rho}(\varphi,\psi) Y_{\rho\alpha}\, \widetilde{Y}^{\sigma\alpha}\biggr],
\label{action}
\end{equation}
where $Y_{\alpha\beta}$ and $\widetilde{Y}^{\alpha\beta}$ are, respectively, the hypercharge field strength and its dual. The symmetric tensors ${\mathcal M}_{\sigma}^{\rho}(\varphi,\psi)$,  ${\mathcal N}_{\sigma}^{\rho}(\varphi,\psi)$  and 
${\mathcal Q} _{\sigma}^{\rho}(\varphi,\psi)$ contain the couplings of the hypercharge either to the inflaton field 
itself (be it for instance $\varphi$) or to some other spectator field (be it for instance $\psi$).  The typical derivative coupling arising in the relativistic theory of Casimir-Polder and Van der Waals interactions \cite{such} is implicitly contained in Eq. (\ref{action}) when  ${\mathcal Q} _{\sigma}^{\rho}(\varphi,\psi)=0$: in this case Eq. (\ref{action}) offers a viable framework for inflationary magnetogenesis  \cite{SUSC1} characterized by unequal electric and magnetic susceptibilities. For immediate convenience the three symmetric tensors appearing in Eq. (\ref{action}) can be parametrized as follows:
\begin{eqnarray}
 {\mathcal M}^{\lambda}_{\,\, \rho} &=& - \frac{\lambda}{16\pi} \delta^{\lambda}_{\,\rho} - \frac{\lambda_{E}(\varphi, \psi)}{16 \pi}\,  \, u^{\lambda} \, u_{\rho}, \qquad  {\mathcal N}^{\lambda}_{\,\, \rho}  = - \frac{\lambda_{B}(\varphi, \psi)}{16 \pi}\, \overline{u}^{\lambda} \, \overline{u}_{\rho},
\nonumber\\
{\mathcal Q}^{\lambda}_{\,\, \rho} &=&  \frac{1}{16 \pi}\,  [\lambda_{1}(\varphi,\psi) \delta^{\lambda}_{\rho} + \lambda_{2}(\varphi,\psi) \, \overline{u}^{\lambda} \, \overline{u}_{\rho}],
\label{POS}
\end{eqnarray}
where $u_{\rho} = \partial_{\rho}\varphi/\sqrt{g^{\alpha\beta} \partial_{\alpha} \varphi \partial_{\beta} \varphi}$ and 
$\overline{u}_{\rho} = \partial_{\rho}\psi/\sqrt{g^{\alpha\beta} \partial_{\alpha} \psi \partial_{\beta} \psi}$ are the normalized 
gradients of the scalar fields.  When all the  coupling functions of Eq. (\ref{POS})  vanish except for $\lambda$, we recover the standard situation where the scalar fields are only coupled to the gauge kinetic term of the hypercharge without Van der Waals interactions. In a conformally flat background geometry\footnote{We focus now on the case where the background metric is conformally flat, i.e. $g_{\mu\nu}(\tau) = a^{2}(\tau) \eta_{\mu\nu}$ where $\eta_{\mu\nu}$ is the Minkowski metric and $\tau$ is the conformal time coordinate.} the gauge field strength can be expressed in terms of the electric and magnetic fields as 
$Y^{i0} = e^{i}/a^2$ and $Y^{ij} = - \epsilon^{ijk} b_{k}/a^2$, the explicit components of ${\mathcal M}_{\rho}^{\sigma}$ and  ${\mathcal N}_{\rho}^{\sigma}$ can be directly obtained from Eqs. (\ref{POS}).

The (comoving) hyperelectric and hypermagnetic are defined, respectively, as
$\vec{B}_{Y} = a^2 \, \sqrt{\Lambda_{B}}\, \vec{b}$ and  $\vec{E}_{Y} = a^2 \, \sqrt{\Lambda_{E}}\, \vec{e}$; note that $\Lambda_{B} = (\lambda + \lambda_{B}/2)$ and $\Lambda_{E} = (\lambda + \lambda_{E}/2)$ can be physically interpreted as the squares of the hypermagnetic and hyperelectric susceptibilities. The evolution equations of $\vec{E}_{Y}$ and $\vec{B}_{Y}$ 
follow directly from Eqs.  (\ref{action})--(\ref{POS}) and they are:
\begin{eqnarray}
&& \vec{\nabla} \times \biggl( \sqrt{\Lambda_{B}} \vec{B}_{Y} \biggr) = \partial_{\tau} \biggl( \sqrt{\Lambda_{E}} \vec{E}_{Y} \biggr) + 4 \pi \vec{J}_{Y} + \partial_{\tau} \biggl[\frac{\Lambda_{BE}}{\sqrt{\Lambda_{B}} } \vec{B}_{Y}\biggr] + \vec{\nabla}\times \biggl[ \frac{\Lambda_{BE}}{\sqrt{\Lambda_{E}}} \vec{E}_{Y} \biggr],
\label{one}\\
&& \vec{\nabla} \times \biggl(\frac{\vec{E}_{Y}}{\sqrt{\Lambda_{E}}}\biggr) + \partial_{\tau} \biggl(\frac{\vec{B}_{Y}}{\sqrt{\Lambda_{B}}}\biggr) =0,
\label{two}\\
&& \vec{\nabla} \cdot \biggl(\frac{\vec{B}_{Y}}{\sqrt{\Lambda_{B}}}\biggr)=0,\qquad \vec{\nabla}\cdot ( \sqrt{\Lambda_{E}}\, \vec{E}_{Y} ) + \vec{\nabla}\cdot \biggl[  \frac{\Lambda_{BE}}{\sqrt{\Lambda_{B}}} \vec{B}_{Y}\biggr]= 4 \pi \rho_{Y},
\label{three}
\end{eqnarray}
where $\Lambda_{BE} = (2 \lambda_{1} + \lambda_{2}/2)$ and $\partial_{\tau}$ denotes the derivative with respect to the conformal time coordinate. The hyperelectric and hypermagnetic couplings are, respectively, $g_{E} = (4\pi/\Lambda_{E})^{1/2}$ and as $g_{B} = (4\pi/\Lambda_{B})^{1/2}$.
When $\Lambda_{BE} \to 0$ and in the absence of sources (i.e. $\vec{J}_{Y} \to 0$ and $\rho_{Y} \to 0$), Eqs. (\ref{one}), (\ref{two}) and (\ref{three}) are invariant under a symmetry\footnote{More specifically
the generalized duality transformation stipulates that under the exchange and inversion of the susceptibilities ($\sqrt{\Lambda}_{E} \to 1/\sqrt{\Lambda_{B}}$ and $\sqrt{\Lambda_{B}} \to 1/\sqrt{\Lambda_{E}}$)  or of the corresponding couplings (i.e.  $g_{E} \to 1/g_{B}$ and $g_{B} \to 1/g_{E}$) Eqs. (\ref{one}), (\ref{two}) and (\ref{three}) maintain the same form provided  the electric and magnetic fields are also exchanged as $\vec{E} \to - \vec{B}$ and  $\vec{B} \to \vec{E}$.} that generalizes the conventional  duality transformation \cite{duality1}. 
While the expression of ${\mathcal M}_{\rho}^{\sigma}$ and ${\mathcal N}_{\rho}^{\sigma}$ may contain 
supplementary terms, a general analysis shows that these terms will simply modify the relation of $\Lambda_{B}$ and $\Lambda_{E}$ to the parameters appearing in the Lagrangian without altering the form of Eqs. (\ref{one}), (\ref{two}) and (\ref{three}).  

The duality symmetry is explicitly broken when $\Lambda_{BE} \neq 0$. In this case the hypermagnetic gyrotropy is amplified from the vacuum fluctuations of the hyperelectric and hypermagnetic fields. To estimate this effect let us consider $\Lambda_{E}$, $\Lambda_{B}$ and $\Lambda_{BE}$ as time dependent but otherwise homogeneous.  The hyperelectric and hypermagnetic field operators can be represented in Fourier space as\footnote{We use the circular polarization 
basis where $\epsilon^{(\pm)}_{i}(\hat{k}) = [ e^{(1)}_{i} \pm i e^{(2)}_{2}]/\sqrt{2}$, and  
$\vec{k} \times \vec{\epsilon}^{\,\,(\pm)}(\hat{k}) = \mp i \, k  \, \vec{\epsilon}^{\,\,(\pm)}(\hat{k})$; $\hat{e}_{1}$, $\hat{e}_{2}$ and $\hat{k}$ are a set of mutually orthgonal unit vectors. In Eqs. (\ref{test3}) and (\ref{test3a}) the sums run over the circular polarizations, i.e. $\alpha=\pm$.}:
\begin{eqnarray}
\hat{B}_{i}^{(Y)}(\vec{p},\eta) &=& - \frac{i}{\sqrt[4]{f}} \sum_{\alpha} \biggl\{ [ \vec{p} \times \vec{\epsilon}^{\,\,(\alpha)}]_{i} \,\, \overline{F}_{p,\,\alpha}(\eta) \, \hat{a}_{\vec{p},\alpha} - [ \vec{p} \times \vec{\epsilon}^{\,\,(\alpha)\, *}]_{i} \,\,\hat{a}^{\dagger}_{-\vec{p},\alpha}  \overline{F}^{*}_{p,\,\alpha}(\eta)\biggr\},
\label{test3}\\
\hat{E}_{i}^{(Y)}(\vec{p},\eta) &=& - \frac{1}{\sqrt[4]{f}} \sum_{\alpha} \biggl\{ \epsilon^{\,\,(\alpha)}_{i} \,\, \overline{G}_{p,\,\alpha}(\eta) \, \hat{a}_{\vec{p},\alpha} +  \epsilon^{\,\,(\alpha)\, *}_{i} \,\,\hat{a}^{\dagger}_{-\vec{p},\alpha}  \overline{G}^{*}_{p,\,\alpha}(\eta)\biggr\};
\label{test3a}
\end{eqnarray}
$\eta$ is an auxiliary time variable defined as $ \sqrt{f} \, d\tau = d\eta$ where  $f = \Lambda_{E}/\Lambda_{B}$. In the limit $f\to 1$ the conformal time variable coincides with $\eta$.  In Eqs. (\ref{test3}) and (\ref{test3a}) $\overline{F}_{p,\,\alpha}(\eta)$ and $\overline{G}_{p,\,\alpha}(\eta)$ (with $\alpha = \pm$) are the mode functions obeying the following evolution equations:
\begin{eqnarray}
\overline{F}_{k,\,\pm}^{\,\,\prime\prime} + \biggl(k^2 - \frac{\sqrt[4]{\Lambda_{E}\, \Lambda_{B}}^{\,\,\prime\prime} }{\sqrt[4]{\Lambda_{E}\, \Lambda_{B}}}\biggr) \overline{F}_{\pm} \pm k\,\frac{\Lambda_{BE}^{\prime}}{\sqrt{\Lambda_{B} \Lambda_{E}}} \overline{F}_{\pm} =0, \qquad \overline{G}_{k,\,\pm} = \overline{F}_{k,\,\pm}^{\,\,\prime} - \frac{\sqrt{\Lambda_{B} \Lambda_{E}}^{\,\,\prime}}{\sqrt{\Lambda_{E}\, \Lambda_{B}}}\,  \overline{F}_{k,\,\pm},
\label{modef}
\end{eqnarray}
where the prime shall denote hereunder a derivation with respect to $\eta$. 

After computing the averages of pairs of field operators at equal times (but different comoving three-momenta) over the initial vacuum state,  the corresponding two-point functions can be readily obtained:
\begin{eqnarray}
\langle \hat{B}^{(Y)}_{i}(\vec{k}, \eta) \, \hat{B}^{(Y)}_{j}(\vec{p},\eta) \rangle &=& \frac{2 \pi^2}{k^3} \delta^{(3)}(\vec{k} + \vec{p}) \biggl[ P_{ij}(\hat{k}) \, P_{B}(k, \eta) 
+ i \, \epsilon_{i j \ell}\, \hat{k}^{\ell} \,P^{(B)}_{{\mathcal G}}(k,\eta) \biggr],
\label{test4}\\
\langle \hat{E}^{(Y)}_{i}(\vec{k},\eta) \, \hat{E}^{(Y)}_{j}(\vec{p},\eta) \rangle &=& \frac{2 \pi^2}{k^3} \delta^{(3)}(\vec{k} + \vec{p}) \biggl[ P_{ij}(\hat{k}) \, P_{E}(k, \eta) 
+ i \, \epsilon_{i j \ell}\, \hat{k}^{\ell} \,P^{(E)}_{{\mathcal G}}(k,\eta) \biggr],
\label{test4a}
\end{eqnarray}
where $P_{ij}(\hat{k}) = (\delta_{ij} - \hat{k}_{i} \hat{k}_{j})$ and $\epsilon_{i j \ell}$ is the three-dimensional Levi-Civita symbol. The hypermagnetic and hyperelectric power spectra appearing in Eqs. (\ref{test4}) and (\ref{test4a}) are:
\begin{equation}
P_{B}(k,\eta) = \frac{k^{5}}{4 \pi^2 \sqrt{f}}\biggl[ | \overline{F}_{k, \, +}|^2 + | \overline{F}_{k, \, -}|^2\biggr],\quad 
P_{E}(k,\eta) = \frac{k^{3}}{4 \pi^2 \sqrt{f}}\biggl[ | \overline{G}_{k, \, +}|^2 + | \overline{G}_{k, \, -}|^2\biggr],
\label{test5a}
\end{equation} 
while $P^{(B)}_{{\mathcal G}}(k,\eta)$ and $P^{(E)}_{{\mathcal G}}(k,\eta)$ are the corresponding gyrotropic contributions:
\begin{equation} 
P^{(B)}_{{\mathcal G}}(k,\eta) = \frac{k^{5}}{4 \pi^2  \sqrt{f}}\biggl[ | \overline{F}_{k, \, -}|^2 - | \overline{F}_{k, \, +}|^2\biggr],\quad 
P^{(E)}_{{\mathcal G}}(k,\eta) = \frac{k^{3}}{4 \pi^2 \sqrt{f}}\biggl[ | \overline{G}_{k, \, -}|^2 - | \overline{G}_{k, \, +}|^2\biggr].
\label{test6a}
\end{equation}

The quasi-de Sitter evolution affects not only 
the energy densities but also the  gyrotropies 
(i.e. ${\mathcal G}^{(B)}= \vec{B}_{Y}\cdot \vec{\nabla} \times \vec{B}_{Y}$ and  
${\mathcal G}^{(E)}= \vec{E}_{Y}\cdot \vec{\nabla} \times \vec{E}_{Y}$).  In the approximation 
of sudden reheating the quasi-de Sitter phase is replaced by the radiation epoch at $\tau_1$ and the value of the comoving conductivity $\sigma_{c}$
depends on the temperature of the plasma so that  $T/\sigma_{c} \simeq {\mathcal O}(2 \alpha^{\prime}) \sim 1/70$ in the limit of high temperatures \cite{mg1,kari}; note that $\alpha^{\prime} = g'^2/(4\pi)$ and $g'\simeq 0.3 $ is the $U(1)_{Y}$ coupling after inflation. Under these conditions the hyperelectric gyrotropy is washed out by finite conductivity effects while the contribution of the hypermagnetic gyrotropy determines the comoving baryon to entropy ratio $\eta_{B} = n_{B}/\varsigma$ \cite{mg1}:
\begin{equation}
\eta_{B}(\vec{x},\tau) = \frac{ 3 \alpha' n_{f}}{8\pi \, H} \biggl(\frac{T}{\sigma_{c}}\biggl) \frac{{\mathcal G}^{(B)}(\vec{x}, \tau)}{a^4 \rho_{crit}},
\label{BAU}
\end{equation}
where $\varsigma= 2 \pi^2 T^3 N_{eff}/45$ is the entropy density of the plasma;
  and  $n_{f}$ is the number of fermionic generations. In what follows $N_{eff}$ shall be fixed to its standard 
 model value (i.e. $106.75$). Equation (\ref{BAU}) holds when the rate of the slowest reactions in the plasma (associated with the right-electrons) is larger than the dilution rate caused by the hypermagnetic field itself \cite{mg1}: at the phase transition 
 the hypermagnetic gyrotropy  is converted back into fermions since the ordinary magnetic fields does not couple to fermions. 

The expectation value of  $\eta_{B}(\vec{x}, \tau_{ew})$, i.e. $\langle \eta_{B}\rangle = \langle \eta_{B}(\vec{x}, \tau_{ew})\rangle$  
depends on the average gyrotropy $\langle {\mathcal G}^{(B)}(\vec{x},\tau_{ew}) \rangle $ and thanks to Eq. (\ref{test4}) the result is:
\begin{equation}
\langle \eta_{B}\rangle = \frac{ 3 \,\alpha^{\prime}\, n_{f}}{4 \pi H_{ew} a^4 \rho_{crit}} \biggl(\frac{T}{\sigma_{c}}\biggr) \int_{0}^{q_{\sigma}} P_{\mathcal G}^{(B)}(q,\tau_{1})\, d q.
\label{BAU2}
\end{equation}
The integral appearing in Eq. (\ref{BAU2}) extends from $q_{ew} \simeq H_{ew} a_{ew}$ 
to $q_{\sigma} \simeq \sqrt{ a_{ew} H_{ew} \sigma_{c}}$ where $q_{\sigma}$ is the hypermagnetic diffusivity scale. If the 
Hubble radius at the electroweak epoch is around $H_{ew}^{-1} \simeq 3$ cm, the diffusivity scale is roughly $10^{-7}\, H_{ew}^{-1}$  
for a typical electroweak temperature ${\mathcal O}(100)$ GeV. The lower extremum of integration can be 
even extrapolated to $0$ since the integrand converges in this limit. 
It is convenient to express the integrand of Eq. (\ref{BAU2}) in terms of the critical fraction of the hypermagnetic energy density multiplied by  $R(q,\tau)$ measuring the  asymmetry of the mode functions in the circular basis: 
\begin{equation}
\langle \eta_{B}\rangle = \frac{3 \,\alpha^{\prime}}{4 \pi} n_{f}  \biggl(\frac{T}{\sigma_{c}}\biggr) \int_{0}^{q_{\sigma}} \frac{d q}{q} R(q, \tau_{1}) \biggl(\frac{q}{q_{ew}}\biggr) \Omega_{B}(q,\tau_{1}), \qquad R(q, \tau) = \frac{|\overline{F}_{-}|^2 - |\overline{F}_{+}|^2}{|\overline{F}_{-}|^2 + |\overline{F}_{+}|^2},
\label{BAU3}
\end{equation}
where we introduced the notation already used in \cite{SUSC1} for $\Omega_{B}(q,\tau) =3 P_{B}(q,\tau)/[4 \pi H^2 a^4 M_{P}^2] $; the analog quantity in the hyperelectric case 
is $\Omega_{E}(q,\tau) = 3 P_{E}(q,\tau)/[4 \pi H^2 a^4 M_{P}^2]$. 

In a mode-independent approach the gauge couplings can always be parametrized as $g_{E}(a) =\overline{g}_{E} (a/a_{1})^{F_{E}}$ and $g_{B}(a) =\overline{g}_{B} (a/a_{1})^{F_{B}}$ for $a \leq a_{1}$ where $F_{E}$ and $F_{B}$ denote the rates of variation in units of the Hubble rate and $a_{1}$ is the scale factor at the end of inflation. In the first quadrant of the $(F_{B},\, F_{E})$ plane where the gauge coupling are both increasing Eq. (\ref{modef}) becomes:
\begin{equation} 
\overline{F}_{k\,\pm}^{\prime\prime} + \biggl[ k^2 - \frac{\sigma^2 - 1/4}{\eta^2} \pm \frac{k \Lambda_{BE}^{\prime}}{\sqrt{\Lambda_{B} \Lambda_{E}}} \biggr] \overline{F}_{k\,\pm}=0,\qquad  \sigma =\frac{1 - 2 F_{E}}{2( 1 + F_{B} - F_{E})}.
\label{F1}
\end{equation}
We now remark that the contribution of $\Lambda_{BE}$ is suppressed in the limit $k \to 0$. Introducing then two numerical factors of order $1$ (i.e. $\gamma$ and $\beta$ in what follows), $\Lambda_{BE}$ can be expressed 
as the sum of of two complementary contributions, namely $\Lambda_{BE} = \gamma \sqrt{\Lambda_{E} \Lambda_{B}} + \beta \eta_{1} \sqrt{\Lambda_{E} \Lambda_{B}}^{\,\,\prime}$. Recalling that $\eta_{1} =\tau_{1}$, if either $\gamma$ or $\beta$ vanish 
separately the leading contribution to $R(q, \tau_{1})$ goes as $q\tau_{1}$ up to logarithmic corrections. If both $\beta$ and $\gamma$ are 
present the contribution of the term containing $\beta$ is always the leading one. Note, incidentally, that 
Eq. (\ref{F1}) can be written in one of the well known forms of the Whittaker's equation (see e.g. \cite{coulomb} for the first applications of this equation to the evolution of the cosmological inhomogeneities). Recalling that the amplified modes satisfy $k \tau_{1} < 1$, we also have that  
$R(q,\tau_{1}) = 2\beta (2 |\sigma| - 1) \,k\tau_{1} \,\ln{k\tau_{1}}$.
We can then estimate the integral over the modes and obtain, from Eq. (\ref{BAU3}), the following rather general result:
\begin{eqnarray}
\langle \eta_{B}\rangle &=& {\mathcal C}_{B}(F_{B}, F_{E}, \beta) \,\epsilon \,{\mathcal A}_{{\mathcal R}} \biggl(\frac{q_{\sigma}}{q_{ew}}\biggr)^{7 - 2 |\sigma|} 
\biggl(\frac{q_{ew}}{q_1}\biggr)^{6 - 2 |\sigma|} [ (7 - 2 |\sigma|) \ln{(q_{\sigma}/q_{1})} -1],
\nonumber\\
{\mathcal C}_{B}(F_{B}, F_{E}, \beta) &=&\beta\,\, \frac{2^{2 (|\sigma|+1)} \Gamma^2(|\sigma|) ( 2 |\sigma|-1)}{3 \pi ( 7 - 2 |\sigma|)^2} \, |1 +F_{B} - F_{E}|^{ 2 |\sigma| -1},
\label{F6}
\end{eqnarray}
where, for $T_{ew} \simeq 10^{2}$ GeV we have that the two previous dimensionless ratios are given, respectively, by $(q_{\sigma}/q_{ew}) =8.43\times 10^{7} (\sigma_{c}/T_{ew})^{1/2} $ and by $(q_{ew}/q_{1}) = 3.79\times 10^{-17}(\epsilon {\mathcal A}_{{\mathcal R}})^{-1/4}$. 
Note that ${\mathcal A}_{{\mathcal R}}$ is the amplitude of the scalar power spectrum at the pivot wavenumber $0.002\,\mathrm{Mpc}^{-1}$ 
and $\epsilon$ is the slow-roll parameter. 

\begin{figure}[!ht]
\centering
\includegraphics[height=7.5cm]{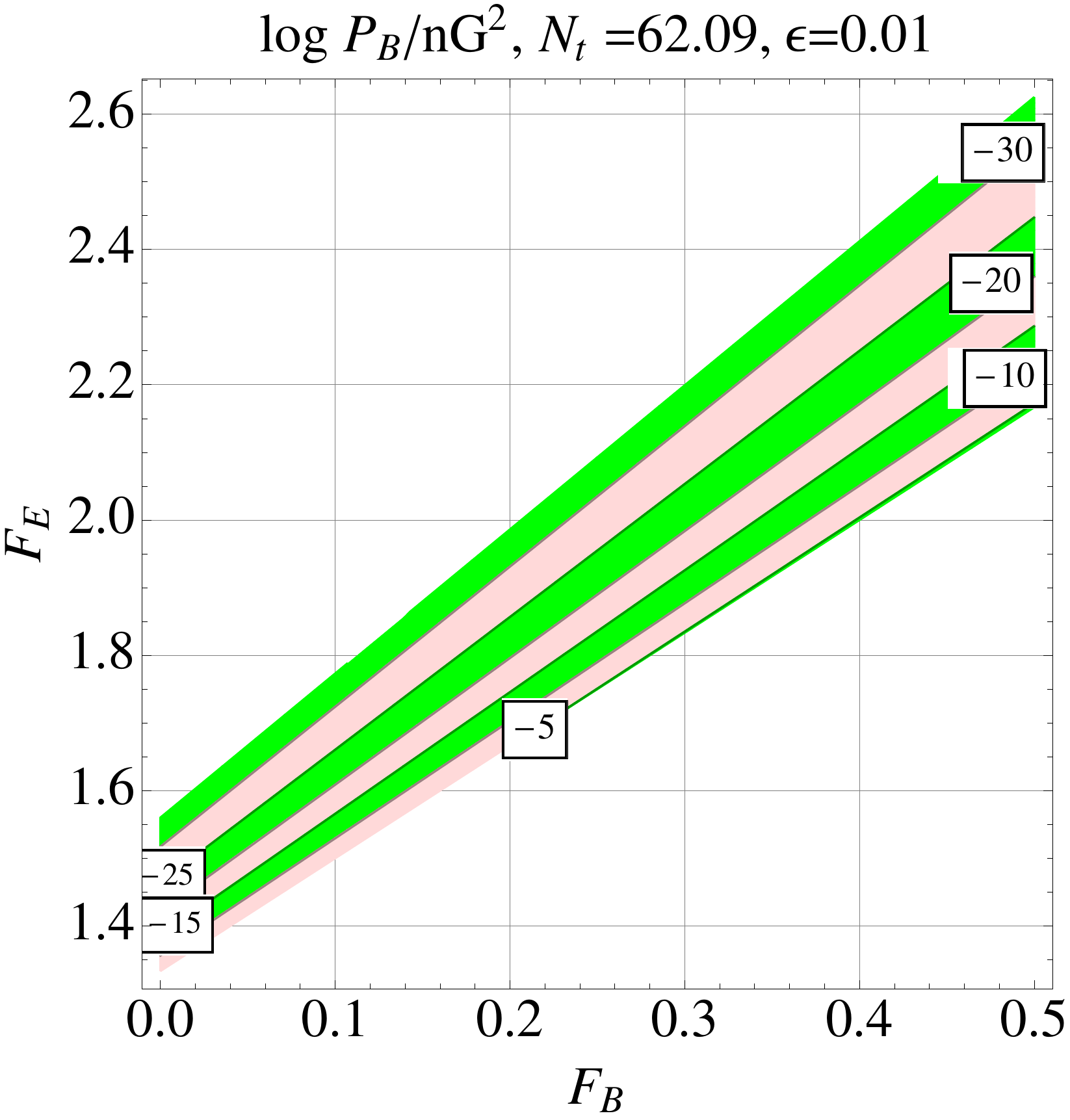}
\includegraphics[height=7.5cm]{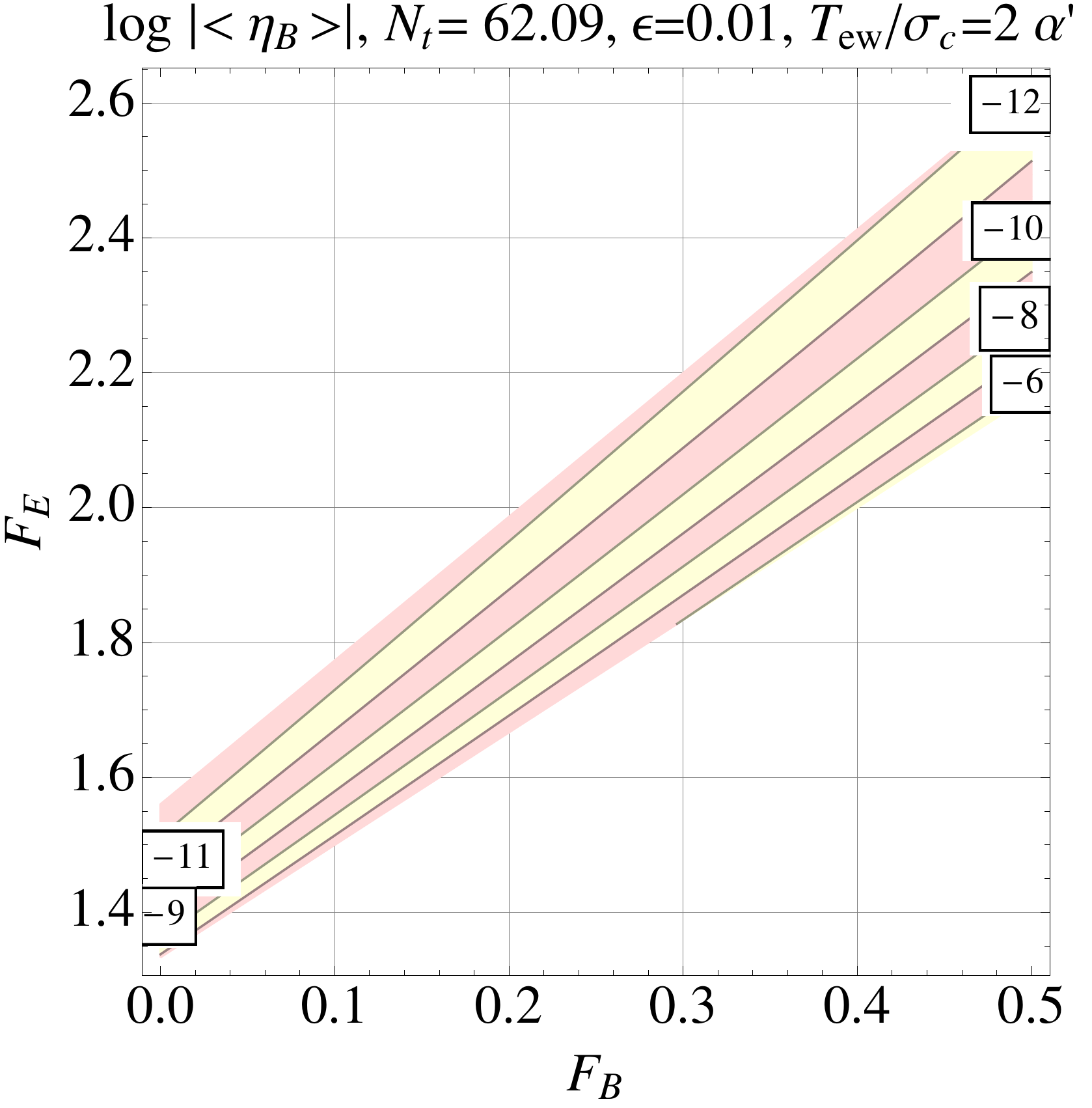}
\caption[a]{In the plot on the left we illustrate the common logarithm of the magnetic power spectrum in the $(F_{B}, F_{E})$ plane; in the plot on the right 
we report the common logarithm of the BAU. In both plots the boundaries of the allowed regions coincide with $1.56 + 2.13 F_{B}$ (upper boundaries) and with 
$5 F_{B}/3 + 4/3$ (lower boundaries). In both plots we took $\beta =1$ since we do not want to tune the pseudo-scalar coupling to a particularly 
small (or a particularly large) value.}
\label{Figure1}      
\end{figure}

The functional dependence of the magnetic spectral index upon $F_{B}$ and $F_{E}$ in the first quadrant of the $(F_{E},\, F_{B})$ plane is given by $n = (5 + 6 F_{B} - 4 F_{E})/(1 + F_{B} - F_{E})$ (the scale-invariant limit corresponds to $n\to 1$, see also \cite{SUSC1}).
For illustration it is interesting to report the magnetic power spectrum at the present time and the BAU at the electroweak epoch in the nearly scale-invariant 
limit for the fiducial choice of the parameters\footnote{Note that the hypercharge field projects onto the electromagnetic field 
through the cosine of the Weinberg angle. Recall also that $\mathrm{nG} = 10^{-9} \mathrm{G}$.} mentioned above:
\begin{eqnarray}
\frac{P_{B}}{\mathrm{nG}^2} &=& 10^{-2.8} \cos^2{\theta_{W}} \biggl(\frac{h_{0}^2 \Omega_{R0}}{4.15\times 10^{-5}}\biggr) \biggl(\frac{{\mathcal A}_{{\mathcal R}}}{2.41\times 10^{-9}}\biggr) \biggl(\frac{\epsilon}{0.01}\biggr) \biggl(\frac{1 + 2 F_{B}}{5} \biggr)^4,
\label{pp}\\
|\langle \eta_{B} \rangle| &=& 1.35 \,\beta \,( \epsilon {\mathcal A}_{{\mathcal R}})^{3/4}\, | 1 - 2 F_{B}|^{4} \biggl[ 18.33 + 0.24 \ln{(\epsilon {\mathcal A}_{{\mathcal R}})}\biggr],
\label{ee}
\end{eqnarray}
implying that $|\langle \eta_{B} \rangle|$ can be as large as  $10^{-7}$. As we shall see from Fig. \ref{Figure1} 
the values of $F_{E}$ and $F_{B}$ (or $\beta$)  can be tuned to obtain a value ${\mathcal O}(10^{-10})$ we stress that larger values are 
phenomenologically safer since, after the electroweak epoch, different physical processes can release entropy 
and further reduce the generated BAU. The scale-invariant limit corresponds, in both plots of Fig. \ref{Figure1}
to the two lower boundaries with equation $F_{E} = 5 F_{B}/3 + 4/3$. 

The physical region of Fig. \ref{Figure1} is therefore located between the scale-invariant limit and the upper boundary determined by the magnetogenesis requirement, i.e. $5 F_{B}/3 + 4/3 \leq F_{E} \leq 1.46 + 1.91 F_{B}$. In this region we have that 
$P_{B}(q_{*}, \tau_{g}) \geq 10^{-22} \mathrm{nG}^2$ where $\tau_{g}$ denotes the time of the protogalactic collapse and $q_{*}$ 
a typical comoving wavenumber ranging between $0.01\, \mathrm{Mpc}^{-1}$ and few $\mathrm{Mpc}^{-1}$ \cite{SUSC1}. This constraint 
can be relaxed down to $10^{-32} \, \mathrm{nG}^2$ (depending on the subsequent protrogalactic evolution) and therefore the allowed region may become 
even wider, i.e. $5 F_{B}/3 + 4/3 \leq F_{E} \leq 1.56 + 2.13 F_{B}$.
The backreaction constraints stipulate that $\Omega_{B}(q,\tau)$ and $\Omega_{E}(q,\tau)$ integrated 
over $d \ln{q}$ must be, at most, $10^{-3}$ and this requirement excludes the other regions in the $(F_{B}, F_{E})$ plane, as it can be 
appreciated from Fig. \ref{Figure1}. Note that in Fig. \ref{Figure1} the total number of efolds $N_{t}$ coincides with $N_{\mathrm{max}}= 63.25 + 0.25 \ln{\epsilon}$ which is the maximal number of efolds which are today accessible to our observations.

In summary the hypermanetic gyrotropy can be amplified from vacuum fluctuations during a quasi-de Sitter stage of expansion when the hyperelectric and the hypermagnetic susceptibilities evolve at different rates. If the gauge couplings unify at the end of inflation there is only one region in the $(F_{E},\, F_{B})$  plane where all the physical requirements (i.e. the  backreaction constraints, the magnetogenesis bounds  and the naturalness of the initial conditions of the scenario) are jointly satisfied.  Since the coupling of hypermagnetic fields to fermions is chiral the produced hypermagnetic gyrotropy may seed the BAU with typical values ranging between $10^{-7}$ and $10^{-10}$ at the electroweak epoch. 
While the present proposal is certainly less conventional than the standard realizations of baryogenesis and leptogenesis, the scope of this investigation has just been to suggest and partially scrutinize an explicit model where the hypermagnetic gyrotropy seeds the BAU in the framework of a consistent magnetogenesis scenario.

\end{document}